\documentclass[prl,twocolumn,showpacs]{revtex4}
\usepackage{amssymb,amsfonts,amsmath,graphicx,bm}
\begin{document}

\title{Tamm surface resonances in very low energy electron scattering from clean metal surfaces}
\date{30 June 2005}
\author{M. N. Read}

\affiliation{School of Physics, University of New South Wales, \\ 
Sydney NSW
2052, Australia}
\email{M.Read@unsw.edu.au}

\begin{abstract}
Very low-energy features which occur in electron reflectivities from clean fcc metal (111) surfaces have been subject to a number of interpretations. Here we analyse the feature near 19.8 eV on Cu(111) at normal incidence and find that it is due to resonant scattering at the rise of the muffin-tin average interstitial potential between atomic layers on approach to the surface from the bulk. This new mechanism corresponds to a Tamm-type surface resonance which is very different in formation to the usual Shockley and Rydberg resonances and explains all features in a systematic way.
\end{abstract}
\pacs{61.14.Hg, 68.37.Nq, 73.20.At}
\maketitle

The copper (111) surface is of special interest because it has a band gap between the Fermi level and vacuum level where a Shockley surface state exists. Adsorbed alkali metals on this surface produce interface states which also lie in this region.  These states form a quasi 2D free-electron gas within a quantum well and it has been suggested that such systems can be used for device application at room temperature \cite{1}. It is therefore important that properties of the clean metal surface are completely determined.  These can be studied by photoemission (PE), inverse photoemission (IPE) spectroscopies and other techniques which give (mostly) below vacuum energy level data. Above-vacuum energy-level data can be found from very low energy electron scattering spectroscopies such as very low energy electron diffraction (VLEED), low energy electron microscopy (LEEM) and target current spectroscopy (TCS).  Both regions can be analyzed theoretically by a layer-by-layer KKR scattering method \cite{2} and properties determined from both energy regions must exhibit a smooth continuation over the whole energy range. This approach effectively provides an expanded data base from which to unravel the complicated surface properties, many of which are expected to have significant energy and momentum variation. Information obtained from the higher energy region also has relevance for analyzing the below-vacuum region.

Here we analyze VLEED and LEEM data on Cu(111). The experimental data at normal incidence has a peak at $\sim 20$ eV which has not been easy to explain.  It has variously been attributed to a band structure effect \cite{3}, a sub-threshold effect \cite{4} and anisotropic inelastic scattering \cite{5}. Similar features are also present for other (111) metal surfaces \cite{3}.  VLEED and LEEM data may have been under-utilized because of the difficulty in systematically accounting for all the features.  This is unfortunate because it is very sensitive to surface scattering potentials and electron self-energy including their energy and momentum variation. It is also more sensitive to the vertical position of surface atoms than LEED which is performed at energies higher than $\sim 40$ eV.

Shockley and Rydberg states and resonances have been detected in below-vacuum PE and IPE experimental data and resonances in above-vacuum VLEED, LEEM and TCS data. Historically surface states (and resonances) have been labeled as Tamm-type and Shockley-type although in some contexts there is no sharp distinction. In a scattering approach, the surface state can be labeled a Shockley type if an essential element in its formation is the shape of the surface barrier in the vacuum beyond the top row of atoms. (They have also been called crystal-induced surface barrier states in this context). Rydberg states arise from the image potential tail of the surface barrier beyond the top row of atoms and are also called image surface barrier states.  In this context a Tamm surface state/resonance could be defined as one which arises from change in the 3D periodicity of the crystal near the surface but is not of the Shockley type as defined above and not due to adsorbed foreign surface atoms. In a scattering picture, the Shockley and Rydberg states/resonances arise because of standing waves forming at some energies in the potential well consisting of the crystal on one side and the rise of the surface barrier to the vacuum on the other. 

With the vacuum emergence of the \{01\},\{10\} beams at 30.7 eV for Cu (111) at normal incidence and an expected crystal inner potential of $\sim 13$ eV \cite{4}, the 20 eV peak lies in the energy range between the crystal and vacuum emergence of the above non-specular beams. In this case scattering of these beams at the surface barrier can produce Shockley resonances. If this situation were the case, the 20 eV peak would give important information about the form of the surface barrier near where it joins to the crystal which has consequences for the determination of other barrier properties. Such a low-energy lying Shockley-type resonance has not been identified to date. 

This possibility has been examined but the 20 eV peak width, height and Bragg-peak separation could not be reproduced by this mechanism for realistic values of electron inelastic scattering of $\sim 2.5$ eV \cite{4,6}. One model came close with the resonance forming a dip in the Bragg peak splitting it into two peaks. This is shown in Fig.~1. However the position of the image origin had to be placed at distances greater than that which is expected theoretically. At realistic positions of the image origin, $z_0$, ($< 4$ a.u. from the centre of the top row of atoms), the well was not wide enough to support a strong resonance only a few eV above its base for electron absorption of $U_{\text{in}} = \;\; \sim 2.5$ eV. The bulk elastic scattering potential used in all the calculations here is due to Moruzzi et al \cite{7}, and it was found to give excellent agreement with band gaps below the vacuum level.

\begin{figure}
\includegraphics[scale=0.45]{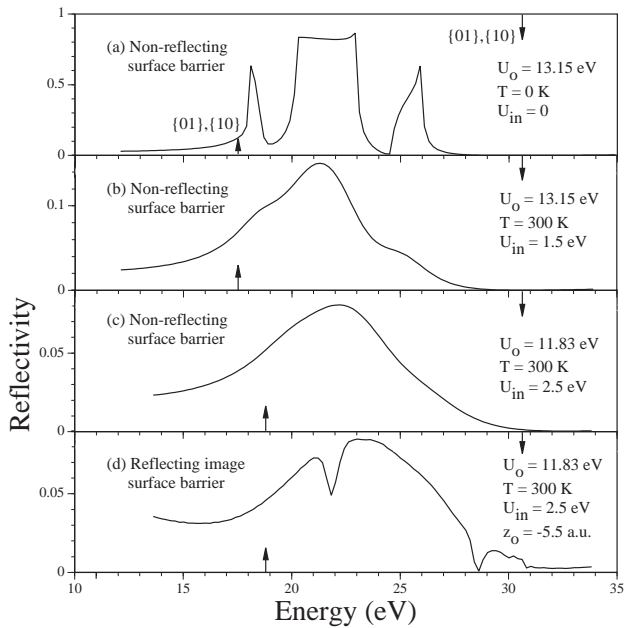}
\caption{Calculated reflectivity of 00 beam on Cu(111) at normal incidence with inner potential $U_{\text{o}}$, uniform inelastic absorption potential $U_{\text{in}}$, crystal temperature {\it T} and Debye temperature 315~K. Frames (a), (b), (c) use a non-reflecting surface barrier. Frame (d) has an image barrier at $z_o = -5.5$ a.u. with saturation to 0.4 eV at $-4.9$ a.u. and barrier inelastic absorption described in Ref.~8 with $\alpha = 1.33$ a.u.  The downward/upward arrows indicate vacuum/crystal emergence of the \{01\},\{10\} beams respectively.}
\label{fig1}
\end{figure}

Another means by which a resonance could occur a few eV above the crystal-emergence energy of the \{01\},\{10\} beams is because of a rise in the muffin-tin average interstitial potential between atomic layers from the bulk value on approach to the surface.  This is illustrated in Fig.~2.  Such changes must occur because of the change in environment of surface atoms.  No features due to this change in the 3D periodicity of the crystal appear to have been identified in low energy electron spectroscopies to date. From a scattering point of view, a potential well is set up where standing waves can be formed between the crystal Bragg pseudo-gap (strong bulk reflection) and the potential rise.  In a scattering picture, this is how a Tamm-type surface resonance is formed. Such a rise in interstitial potential could also produce Tamm surface states below the vacuum level in surface-projected bulk band gaps and resonances outside these gaps. The point is that potential variations detected in above-vacuum spectroscopies (like VLEED, LEEM) also have analogies for below-vacuum spectroscopies (PE, IPE etc). 

\begin{figure}
\includegraphics[scale=0.4]{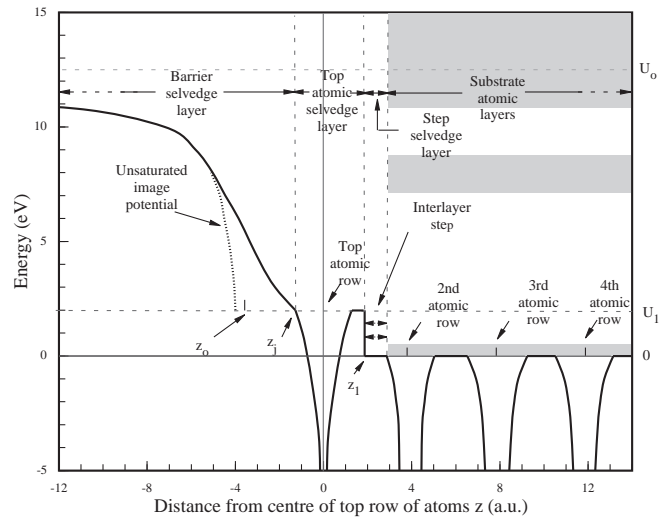}
\caption{Schematic potential energy at the Cu(111) surface in a direction perpendicular to the surface. Symbols are explained in the text except for $z_j$, the jellium discontinuity. Dashed horizontal lines with double-ended arrows near $z_1$ indicate electron scattering which gives rise to standing waves and the feature near 20 eV in experimental data. Unshaded regions in the substrate represent surface-projected bulk-band gaps.}
\label{fig2}
\end{figure}

\begin{figure*}
\includegraphics[scale=0.45]{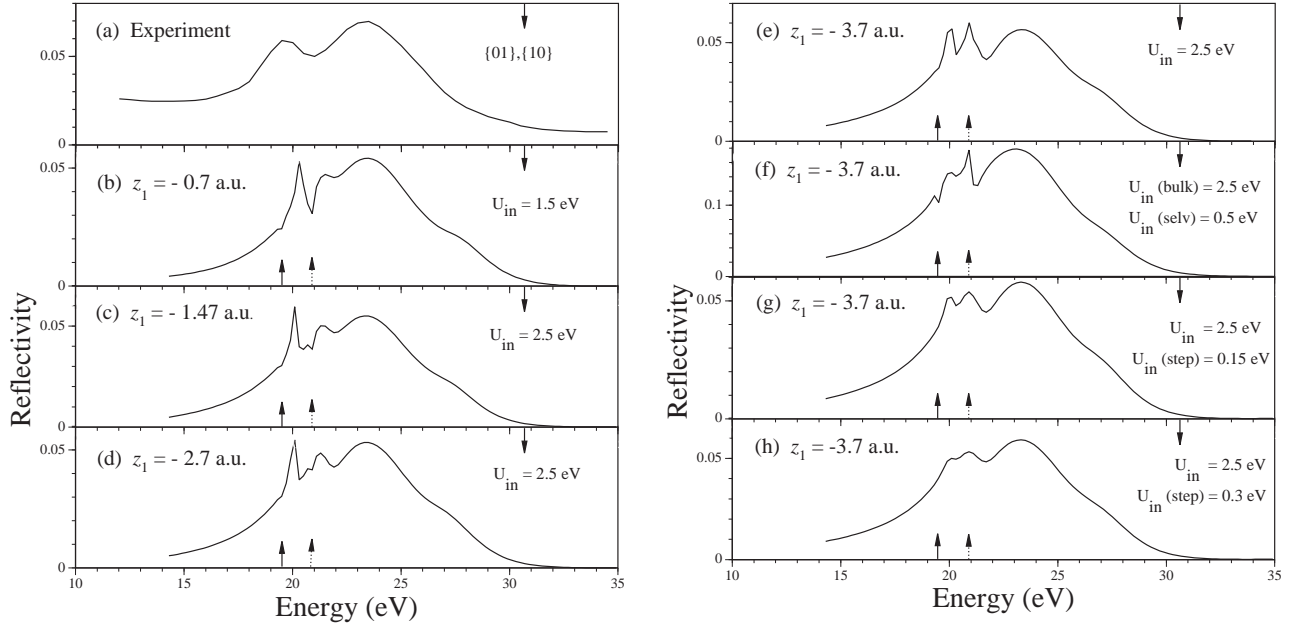}
\caption{Reflectivity of 00 beam on Cu(111) at normal incidence. Frame (a) is the experimental result of Ref.~5 with energy step 0.5 eV. Frames (b) $\rightarrow$ (h) are calculated profiles with $U_o = 11.15$ eV and $U_{\text{in}} = 2.5$ eV in all layers except frame (f) which has $U_{\text{in}} = 0.5$ eV in the top atomic layer. Symbols are described in Fig.~1. In all cases an interlayer step of height $U_1 = 1.4$ eV is placed between the top layer and the next layer at positions $z_1$ from the centre of the second row of atoms at the surface, as indicated on the diagrams. The surface barrier height is 9.75 eV, the crystal temperature is 300~K and Debye temperature 315~K. Downward/upward arrows are the same as in Fig.~1. Dashed upward arrows indicate crystal emergence energy of \{01\},\{10\} beams in the top atomic layer. Frames (g) and (h) have inelastic absorption at the interlayer step as described in the text.}
\label{fig3}
\end{figure*}

To examine this case, calculations were performed with an interlayer potential rise of height $U_1 = 0.1029$ Ry = 1.4 eV  with respect to the bulk muffin-tin zero (crystal inner potential) placed {\it between} the top layer of atoms and the atomic layer below.  This is illustrated schematically in Fig.~2. The top layer was followed by a surface barrier of height $U_2 = 9.75$ eV with respect to $U_1$. It was found that the shape of the surface barrier was not important here since it does not give rise to any barrier features of the Shockley or Rydberg type when the image origin is placed at any realistic position from the centre of the top layer of atoms and realistic crystal electron absorption of $>2.3$ eV is used.  The layer KKR method for calculating LEED reflectivities of McRae \cite{9} was used where the crystal substrate is terminated at $z = 0$ and $z$ is the distance perpendicular to the surface and positive directed into the crystal. Cu(111) has a lattice constant of 6.8309 a.u. with 3 layers in the repeating unit. The top layer of atoms is taken as part of the selvedge and for Cu(111) is placed at $z = -3.9438$ a.u. and translated with respect to the first substrate layer which has an atom at the origin by $x = 1.6102$ and $y = 3.2201$ a.u. The interlayer rise in potential is modeled in these calculations by a simple step potential placed between $z = 0$ and -3.9438 a.u. From higher energy LEED analyses on Cu(111) the surface atomic layers have not been found to deviated significantly from their bulk positions \cite{4}. From the position $z = -3.9438$ a.u. there is a further rise in potential representing the surface barrier of height $U_2$. The total rise in potential from the bulk muffin-tin zero to vacuum level is $U_1 + U_2 = U_0 = 11.15$ eV in the calculations shown here.

Wave vectors of the wave functions corresponding to beam $v$ in the three regions bulk, selvedge and vacuum are represented by $\bm{k}_v^{\pm}$, $\overline{\bm{k}}\,\!_v^{\pm}$ and $\bm{K}_v^{\pm}$ respectively after the notation of McRae \cite{9}. For the interlayer step, the reflection and transmission coefficients are
\begin{subequations}
\begin{eqnarray}
\rho_S(\bm{k}_v^+ \bm{k}_v^-) & = & \frac{(k_v^{\perp} - \overline{k}\,\!_{v}^{\perp})}{(k_v^{\perp} + 
\overline{k}\,\!_v^{\perp})} \exp[-2iz_1k_v^{\perp}]. \\
\tau_S(\overline{\bm{k}}\,\!_v^- \bm{k}_v^-) & = & \frac{2 k_v^{\perp}}{(k_v^{\perp} + \overline{k}\,\!_v^{\perp})} \exp[iz_1(\overline{k}\,\!_v^{\perp} - k_v^{\perp})]. \\
\rho_S(\overline{\bm{k}}\,\!_v^- \overline{\bm{k}}\,\!_v^+) & = & \frac{(\overline{k}\,\!_v^{\perp} - k_v^{\perp})}{(\overline{k}\,\!_v^{\perp} + k_v^{\perp})} \exp[2iz_1\overline{k}\,\!_v^{\perp}]. \\
\tau_S(\bm{k}_v^+ \overline{\bm{k}}\,\!_v^+ ) & = & \frac{2 \overline{k}\,\!_v^{\perp}}{(\overline{k}\,\!_v^{\perp} + k_v^{\perp})} \exp[iz_1(- k_v^{\perp} + \overline{k}\,\!_v^{\perp})].
\end{eqnarray}
\end{subequations}
where $k_v^{\perp}$, $\overline{k}$$_{v}^{\perp}$ are the perpendicular components of the corresponding  wave vectors and $z_1$ is the position (with respect to the origin) of the potential step. This sharp step overestimates above-step reflections but the aim is to keep the model as simple as possible at this stage and easily reproducible by others. The $\rho_S$ and $\tau_S$ were used to calculate a transfer matrix $\bm{Q}_2$ for this step scattering layer according to Eq.~(14) of McRae \cite{9}.  Similarly transfer matrices for the top atomic selvedge layer $\bm{Q}_1$ and for the surface barrier layer $\bm{Q}_0$ were calculated, giving a selvedge scattering matrix $\bm{X}$ where 
\begin{equation}
\bm{X} = \bm{Q}_2 \cdot \bm{Q}_1 \cdot \bm{Q}_0 .
\end{equation}
The selvedge scattering matrix $\bm{X}$ was combined with the bulk crystal scattering matrix $\bm{M}$ using Eq.~(35) of McRae \cite{9}. In order to keep the model as uncomplicated as possible, the phase shifts for the top atomic selvedge layer were kept the same as the bulk. The bulk isotropic inelastic scattering potential $U_{\text{in}}$(bulk) was 2.5 eV and for the selvedge atom layer $U_{\text{in}}$(selv) was chosen from 0.5 to 2.5 eV. Different heights $U_1$ of the interlayer step (and barrier height $U_2$) were tried.

Fig.~3 shows the results of our calculation for interlayer step positions from $z_1$ = -0.7 to -3.7 a.u. with $U_1$ = 1.4 eV.  In frames (b) to (d), no inelastic scattering was included for the step layer, i.e. the normal components of the wave vectors in Eq.~(1) had no imaginary components. This of course overestimates the effects but allows one to see the possibilities. Firstly, the value of inelastic scattering in the selvedge top layer, $U_{\text{in}}$(selv), has some effect on the fine structure features but for this simple model, optimization of such parameters was not considered appropriate at this stage. In comparison with the reflectivity without the interlayer step as seen in frame (a) of Fig.~1, we see that a peak appears near 20 eV and moves down in energy from 20.3 to 20 eV as $|z_1|$ increases. The six \{01\},\{10\} beams emerge in the bulk crystal layers at 19.5 eV and in the selvedge top atomic layer at 20.9 eV  (with respect to the vacuum level).  The \{01\},\{10\} beams from the bulk are propagating and incident on the step in a direction towards the surface in the energy range above 19.5 eV. The \{01\},\{10\} beams in the top atomic layer are incident on the step in a direction away from the surface and are evanescent until 20.9 eV when they become propagating. Multiple reflections and transmissions in both directions give rise to the \{01\},\{10\} beams becoming incident on the bulk atomic layers and selvedge atomic layers with phases and amplitudes different from what occurs without the step. At certain energies, standing waves may form which correspond, in a scattering picture, to what historically has been called Tamm surface resonances. The non-specular scattering at atomic layers leaks some flux into the 00 beam. Above step-height reflections may be exaggerated here but some will also occur for a smoother interlayer potential rise. 

In this model, inelastic scattering in the whole volume of the selvedge and substrate regions already has been taken into account in the calculation of the bulk and atomic layer scattering matrices. Also inelastic scattering in the top bulk layer is likely to be different from that for lower positioned layers.  It is difficult for a model based on the analytical reflections coefficients in Eq.~(1) to ascribe appropriate inelastic scattering from events associated with the step only.  Changes in the isotropic-layer inelastic scattering potentials on approach to the surface are unknown quantities at this stage.  Hence for this model we have reduced the intensity of the interlayer step features somewhat arbitrarily through the wave vectors in Eq.~(1) by adding an inelastic scattering potential, $U_{\text{in}}$(step), of 0.3 eV and 0.15 eV. This simulates how the features change with inelastic scattering.  We see in frame (e) of Fig.~3 that the structure appears to be of the same form as seen in the experimental data including its energy width and separation from the Bragg peak. Resolution of any fine structure features would also be lost in most experimental set-ups because of beam width and energy broadening. With exaggerated above-step features the model used here would not be expected to exactly reproduce the experimental result, but the origin of the feature can be determined.

In conclusion we have found the following: (a) For normal incidence electron reflectivity from Cu(111), a peak near 20 eV with the given experimental width, height and Bragg-peak separation could not be produced from a surface barrier with image origin, $z_0$, located in a realistic position and for inelastic scattering potential, $U_{\text{in}}$, of realistic value; the peak is not due to a Shockley-type surface barrier resonance.
(b) The ~20 eV peak is due to resonant scattering at the rise in the average interstitial potential between atomic layers on approach to the surface. It is therefore a Tamm-type surface resonance. This type of surface resonance has not been identified before in VLEED, LEEM etc. It is also likely to account for similar features in other clean metal (111) surfaces.
(c) The interlayer rise in potential may only produce surface resonances when they fall in the energy range of a surface-projected bulk band pseudo-gap (Bragg peak) in VLEED and related spectroscopies.
(d) Even when the condition in (c) above is not fulfilled, the rise in potential before the top atomic layer will change the value of the height of the surface barrier. This value affects the determination of barrier features such as image origin position, saturation etc, for both below vacuum level spectroscopies (PE, IPE) and those above (VLEED, LEEM, TCS).
(e) All of the features in VLEED (0 - 40 eV) may now be accounted for in a systematic way if, (i) scattering at a realistic image surface barrier, and (ii) scattering at the interlayer rise in interstitial muffin-tin potential between surface layers is included in the theoretical calculation of reflectivities. This would make VLEED and LEEM  more widely used tools for unravelling surface properties in the future.

\end{document}